\documentclass{elsart}

\usepackage{epsfig}

\usepackage{amssymb}

\begin{document}

\begin{frontmatter}



\title{New experimental limit on the Pauli
 Exclusion Principle violation by electrons}

\vspace{-5mm}
\begin{center}
\large{The VIP Collaboration:}
\end{center}
\vspace{-2mm}
\author[Frascati]{S. Bartalucci},
\author[Frascati]{S. Bertolucci},
\author[Frascati,Bucharest]{M. Bragadireanu},
\author[Vienna]{M. Cargnelli},
\author[Frascati]{M. Catitti},
\author[Frascati]{C. Curceanu (Petrascu)\corauthref{cor}},
\corauth[cor]{Corresponding author. Tel: +39 0694032321, Fax: +39 0694032559}
\ead{Catalina.Petrascu@lnf.infn.it}
\author[Frascati]{S. Di Matteo},
\author[Neuchatel]{J.-P. Egger},
\author[Frascati]{C. Guaraldo},
\author[Frascati,Bucharest]{M. Iliescu},
\author[Vienna]{T. Ishiwatari},
\author[GranSasso]{M. Laubenstein},
\author[Vienna]{J. Marton},
\author[Trieste]{E. Milotti},
\author[Frascati]{D. Pietreanu},
\author[Bucharest]{T. Ponta},
\author[Frascati,Bucharest]{D.L. Sirghi},
\author[Frascati,Bucharest]{F. Sirghi},
\author[Frascati]{L. Sperandio},
\author[Vienna]{E. Widmann},
\author[Vienna]{J. Zmeskal}

\address[Frascati]{INFN, Laboratori Nazionali di Frascati, C. P. 13, Via E. Fermi 40, I-00044, Frascati (Roma), Italy}

\address[Bucharest]{'Horia Hulubei' National Institute of Physics and
 Nuclear Engineering, Str. Atomistilor no. 407, P.O. Box MG-6, 
Bucharest - Magurele, Romania}

\address[Vienna]{ Stefan Meyer Institute for Subatomic Physics,
 Boltzmanngasse 3, A-1090 Vienna, Austria}

\address[Neuchatel]{Institute de Physique, Universit\'e de Neuch\^atel,1 rue A. -L. Breguet, CH-2000 Neuch\^atel, Switzerland}

\address[GranSasso]{Laboratori Nazionali del Gran Sasso, 
S.S. 17/bis, I-67010 Assergi (AQ), Italy}

\address[Trieste]{ Dipartimento di Fisica, Universit\`{a} di Trieste and INFN-- Sezione di Trieste, Via Valerio, 2, I-34127 Trieste, Italy}



\begin{abstract}

The Pauli Exclusion Principle (PEP) is one of the basic principles of modern 
physics and, even if there are no compelling 
reasons to doubt  its validity, it is still debated today because an intuitive,  
elementary explanation is still missing, and because of its unique  
stand among the basic symmetries of physics.
The present paper reports a new limit on the probability 
that PEP is violated by electrons, in a search for a 
shifted K$_\alpha$ line in copper: 
the presence of this line in the soft X-ray copper 
fluorescence would signal a transition to a ground state 
already occupied by 2 electrons. 
The obtained value, $\frac{1}{2} \beta^{2} \leq  4.5\times 10^{-28}$, improves 
the existing limit 
by almost two orders of magnitude.

\end{abstract}

\end{frontmatter}


\section{Introduction}

The Pauli Exclusion Principle (PEP) is a 
consequence 
of the  spin-statistics connection \cite{Pauli} and 
plays a fundamental role in our understanding of many physical
and chemical phenomena, from the periodic table of elements, 
to the electric conductivity in metals,
to the degeneracy pressure, which makes white dwarfs and neutron stars stable,
just to cite few ones.
Although the principle has been  spectacularly confirmed by 
the number and accuracy of its predictions, 
its foundation lies deep in the structure of quantum field 
theory and has defied all attempts to produce a simple proof, 
as nicely stressed by  R. Feynman \cite{Feynman}.

Given its basic standing in quantum theory, it seems appropriate 
to carry out precise tests of the PEP validity  and, indeed, in the last 
fifty years, several experiments have been performed to search for
possible small 
violations \cite{Bernabei,Borexino,HH,NEMO,Nolte,Tsi}. 
Often, these experiments were born as by-products of experiments
with a different objective (e.g. dark matter searches, proton decay, etc.. ), 
and most of the recent limits on the validity of PEP have been obtained 
for nuclei or nucleons.
 
Concerning the violation of 
PEP for electrons, Greenberg and Mohapatra \cite{GM:Gre}  
examined all experimental 
data which could be related, directly or indirectly, to PEP, up to 1987. 
In their analysis they concluded  that the 
probability that a  new electron added to an antisymmetric 
collection of N electrons might
form a mixed symmetry state rather than a totally antisymmetric state 
is $\leq 10^{-9}$. 
In 1988, Ramberg and Snow \cite{RS:RAM} drastically improved this limit with 
a dedicated experiment, searching for 
anomalous X-ray transitions, that would point to a small violation of PEP in  
a copper conductor. 
The result of the experiment was a  probability $\leq 1.7\times10^{-26}$ 
that a new electron  circulating in the 
conductor would form a mixed symmetry state with the already present copper
electrons. 

We have set up an improved version of the  Ramberg and Snow experiment,
with a higher sensitivity apparatus \cite{VIPproposal}. 
Our final aim is to lower the PEP violation limit for electrons
by at least 4 orders of magnitude,
 by using high resolution Charge-Coupled  Devices (CCD)
as soft X-rays detectors \cite{CCD},
 and decreasing the effect of background 
by a careful choice of the  materials and sheltering the apparatus 
in an underground laboratory.  

In the next sections we describe the experimental setup,
the outcome of a preliminary measurement performed 
in the Frascati National Laboratories (LNF) of INFN in 2005, 
along with a brief discussion on the results and 
the foreseen future improvements in the Gran Sasso National Laboratory (LNGS) of
INFN.

\section{The VIP experiment}

The idea of the VIP ({\underline VI}olation of the {\underline P}auli 
Exclusion Principle)
experiment was originated by the 
availability of the DEAR (DA$\Phi$NE Exotic Atom Research) 
setup, 
after it had successfully completed its program at the 
DA$\Phi$NE collider  at LNF-INFN \cite{DEAR}. 
DEAR  used Charge-Coupled Devices (CCD) as detectors in 
order to measure exotic atoms (kaonic nitrogen and 
kaonic hydrogen) X-ray transitions.
CCD's are almost ideal detectors for X-rays measurement, due to their excellent
background rejection capability, based on pattern recognition, and
to their good energy resolution (320 eV FWHM at 8 keV in the present measurement).

\subsection{Experimental method}

The experimental method, originally described in \cite{RS:RAM},
 consists in the introduction of new electrons into a 
copper strip, by circulating a current,
 and in the search for X rays resulting from the $2p \rightarrow 1s$ 
anomalous radiative transition  that occurs if one 
of the new electrons is captured by a copper atom and cascades down 
to the 1s state already filled by 
two electrons of opposite spin.
The energy of this 
transition would differ from the normal K$\alpha$ transition by about 300 eV 
(7.729 keV instead of 8.040 keV) \cite{PI:Pau}, providing an unambiguous signal 
of the PEP violation.  
The measurement alternates periods 
without current 
in the copper strip, 
in order to evaluate the X-ray background in conditions where
no PEP violating transitions are expected to occur, with periods in 
which current flows in the conductor, thus providing ``fresh'' electrons,
which might possibly violate PEP. The fact that no PEP violating transitions are
expected to be present in the measurement without current  is related to
the consideration that any initial conduction electron in the copper that
was in a mixed symmetry state with respect to the other copper electrons,
 would have already cascaded down to the $1s$ state and would therefore 
 be irrelevant for the present experiment.
The rather straightforward analysis  consists in the evaluation 
of the statistical significance
of the  normalized subtraction of the two 
spectra, with and without current,  in the energy region where 
the PEP violating transition is expected.

\subsection{The VIP setup}

The VIP setup  consists of a copper cylinder, 
4.5 cm in radius, 50 $\mu$m thick, 8.8 cm high, surrounded by 
16 equally spaced CCD's \cite{xx:EEV}.
The CCD's are at a distance of 2.3 cm from the copper 
cylinder, grouped in units of two chips vertically positioned.
The setup is shown in Fig. 1.
The chamber is kept at high vacuum to  
minimize X-ray absorption and to avoid condensation on the cold  
surfaces. The copper target (the copper strip where the current flows  
and new electrons are injected from the power supply) is at the  
bottom of the setup. The CCD's surround the  
target and are supported by cooling fingers that start from the  
cooling heads in the upper part of the chamber. The CCD readout  
electronics is just behind the cooling fingers; the  
signals are sent to amplifiers on the top of the chamber. The amplified  
signals are read out by ADC  boards in a data acquisition  
computer.

\begin{figure}
\begin{center}
\mbox{\epsfysize=110mm\epsffile{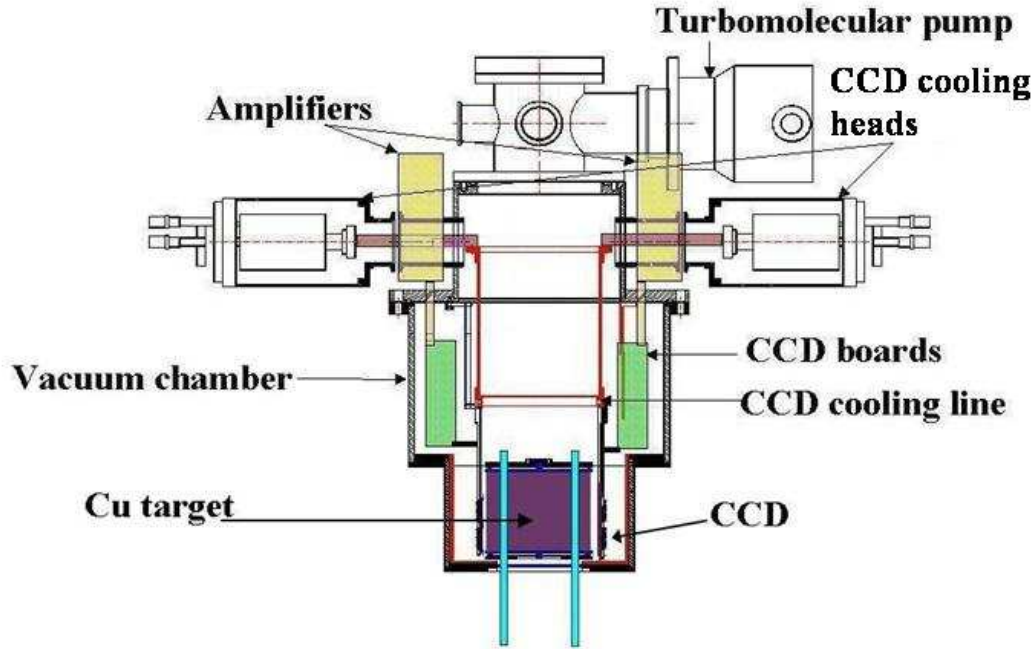}}
\caption{The VIP setup. All elements af the setup are identified in the figure.}
\label{Ponline.EPS}
\vspace{0.5cm}
\end{center}
\end{figure}

More details on the CCD-55 performance, as well as on the
analysis method used to reject background, can be found in \cite{NIM}

\subsection{Measurements}

The measurements reported in this paper have been
performed in the period
 21 November - 13 December 2005, at
the Frascati National Laboratories of INFN, Italy. 

Two types of measurements were performed:
\begin{itemize}
\item{} 14510 minutes (about 10 days)
of measurements with a 40 A current circulating
in the copper target;
\item{} 14510 minutes of measurements without circulating
current,
\end{itemize}
\noindent
where CCD's were read-out every 10 minutes.

The two resulting X-ray spectra are shown in Figure 2 a), with circulating
current,  and b), without current.
The spectra refer to 14 CCD's (out of 16),
 due to noise problems in the remaining 2.

\begin{figure}
\begin{center}
\mbox{\epsfysize=78mm\epsffile{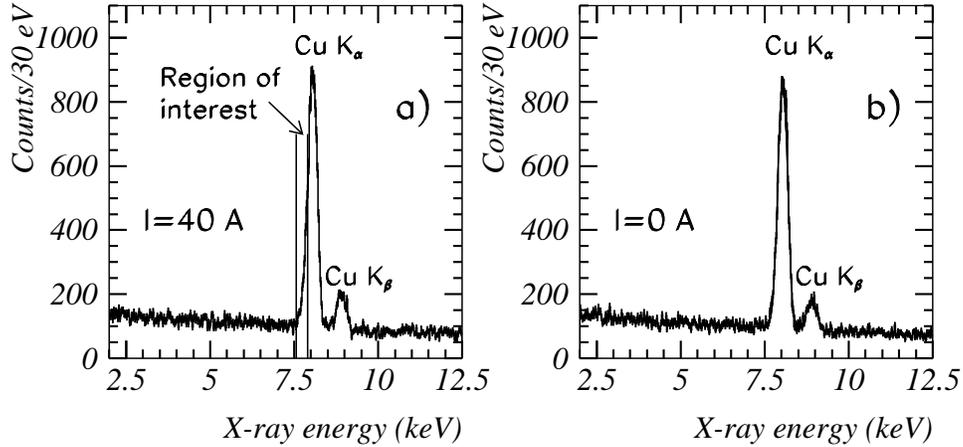}}
\caption{Energy spectra for the VIP measurements:
a) with current (I=40 A); b) without current (I=0).}
\label{P1fullrange.EPS}
\end{center}
\end{figure}

\section{PEP-violating X-ray spectrum}

In order to obtain the number of  X-rays due to the possible PEP
violating transitions, the spectrum without current was subtracted from the
one with current.

The resulting subtracted spectrum is shown in Figure 3 a) (whole energy scale) and
b) (a zoom on the region of interest).
The region of interest, from 7.564 to 7.894 keV, is defined by the
 CCD energy resolution (320 eV FWHM) at the $K_\alpha$
copper transition (8.04 keV), with an additional uncertainty of 10 eV, to
account for the theoretical uncertainty in the calculation of the
PEP violating transition energy.
The numbers of X rays in the region of interest were:
\begin{itemize}
\item{} at I=40 A: N$_X = 2721 \pm 52$;
\item{} for  I=0 A: N$_X =  2742 \pm 52$;
\item{} for the subtracted spectrum:
$\Delta N_X=-21 \pm 73$.
\end{itemize}

\begin{figure}
\begin{center}
\mbox{\epsfysize=78mm\epsffile{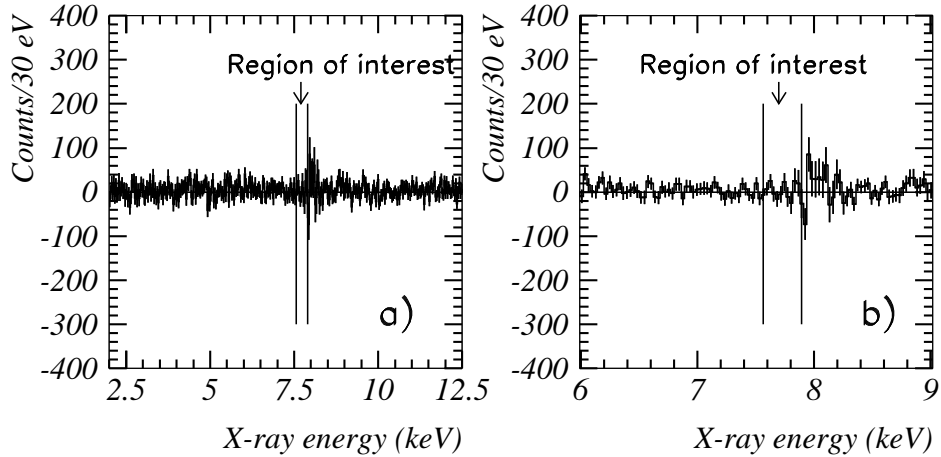}}
\caption{The subtracted spectrum: current minus no-current, giving the
limit on PEP violation for electrons: a) whole energy range; b) expanded view 
in the region of interest (7.564 - 7.894 keV).  
No evidence for a peak in the  region of interest is found.
}
\vspace{0.5 cm}
\label{P18keV.EPS}
\end{center}
\end{figure}

\subsection{Determination of the PEP violation probability limit}

For the parametrization of the results in a Pauli principle violating theory,
we use the notation of  Ignatiev and Kuzmin  \cite{IK}, which has 
been incorporated in the paper of Greenberg and Mohapatra \cite{GM:Gre}: 
even though the model of Ignatiev and Kuzmin has been later shown 
to be incompatible with quantum field theory \cite{Govorkov}, the 
parameter $\beta$ that measures the degree of PEP violation 
has stuck and is still found in the literature, also because it is 
easy to show that it is related to the parameter $q$ of quon theory, by the
relation:
$(1+q)/2 = \beta ^2/2$ \cite{green} (in quon theory, $-1 \le q \le 1$, 
where $q=-1$  corresponds to fermions and $q=1$ corresponds to bosons, 
so that here  $q$ must be close to -1 and $(1+q)/2$ must be very small, 
because we   are dealing with electrons).
Moreover, we used this parametrization for an easy comparison of our results
with the previous Ramberg and Snow ones \cite{RS:RAM}, since the same has
been used in that paper.
In \cite{IK} a pair of electrons in a 
mixed symmetry state has the probability $\beta^2/2$ for 
the symmetric component and $(1-\beta^2/2)$ for the usual antysymmetric one.
The parameter $\beta^{2}/2$ is related, then, to the probability that an 
electron violates PEP (see also  \cite{ok:Oku} for further details).
To determine 
the experimental limit on $\beta^{2}/2$ from our data, we 
used the same arguments of Ramberg and Snow, to compare the results.
The number of electrons that pass through the conductor, 
which are new for this conductor, is:
\begin{equation}
N_{new}= (1/e) \Sigma I \Delta t. 
\end{equation}
\noindent
where $e$ is the electron electric charge, $I$ is the current intensity and
$\Delta t$ represents the time duration of the measurement with current on.
Each new electron will undergo a  large number of 
scattering processes on the atoms of the copper lattice. The minimum number of 
these internal scattering processes per electron, defined as $N_{int}$, 
is of order $D/\mu$, where $D$ is the length 
of the copper electrode (8.8 cm in our case)
and $\mu$ is the mean free path 
of electrons in copper. The latter parameter is obtained from the 
resistivity of the metal. We assume that the capture probability 
(aside from  the factor $\sim \beta^{2}$/2) is greater than $\frac{1}{10}$ 
of the scattering probability. 

The acceptance of the 14  CCD detectors and the probability 
that an X ray of  about 7.6 keV, the energy of the possible anomalous
 transition generated in the copper target, is
not absorbed inside the copper itself, were evaluated by a 
Monte Carlo simulation of the VIP setup,  based on GEANT 3.21. 
This probability turns out to be 2.1\%. 
Moreover, a CCD efficiency equal to 48\% for a 
7.6 keV X ray was considered. All these factors built up 
 the so called  $geometric ~factor$ ($\sim 1$\%).

The number of X rays generated in the PEP violating transition,
$\Delta N_X$, is then related to the $\beta^2/2$ parameter by:

\begin{eqnarray}
\Delta N_{X} & \geq & \frac{1}{2}\beta^{2}N_{new}\frac{1}{10}N_{int}\times 
(geometric~ 
factor)\\
& &  \nonumber \\ 
      & = & \frac{\beta^{2} (\Sigma I \Delta t) D}{e \mu } 
 \frac{1}{20}\times (geometric~ 
factor)
      \nonumber
\end{eqnarray}

Then, for $\Sigma I \Delta t = 34.824 \times 10^{6}$ C, $D = 8.8$ 
cm, $\mu=3.9\times 10^{-6}$ cm, 
 $e=1.602 \times 10^{-19}$ C, we get

\begin{eqnarray}
\Delta N_{X}\geq 4.9  \times 10^{29} \times \frac{\beta^{2}}{2}.
\end{eqnarray}

The difference of events between the measurements with and 
without current, reported in the previous Section, is
$\Delta N_X= -21 \pm 73$. 
Taking as a limit of observation three standard deviations, 
we get for the PEP violating parameter:

\begin{eqnarray}
\frac{\beta^{2}}{2} \leq \frac{ 3\times 73}{4.9\times 10^{29}} = 4.5
\times 10^{-28} ~at \; 99.7 \% \; CL.
\end{eqnarray}

We can interpret this as a limit on the probability of PEP violating 
interactions between external electrons and copper atoms: 
 $\frac{1}{2} \beta^{2} \leq  4.5\times 10^{-28}$. We have thus  
improved the limit obtained by Ramberg and Snow by a factor about 40.

\section{Conclusions and perspectives}

The paper reports a new measurement of the PEP violation limit for electrons,
performed by the VIP Collaboration at LNF-INFN.
The search of a tiny violation was based on a measurement of PEP
violating X-ray transitions in copper, under a circulating 40 A current.
A new limit for the PEP violation for electrons was found:
$\frac{1}{2} \beta^{2} \leq  4.5\times 10^{-28}$, lowering
by almost two orders of magnitude the previous one \cite{RS:RAM}.

We shall soon repeat the measurement
in the Gran Sasso--INFN underground laboratory, at higher integrated currents.
From preliminary tests, it appears that the X-ray background in the
 LNGS environment is a factor 10-100 lower than in the Frascati Laboratories.
A VIP measurement of two years (one with current, one without current)
at LNGS, to start in
Spring 2006, will then bring the limit on PEP violation for electrons
 into  the 10$^{-30}$-10$^{-31}$ region, which is
of particular interest \cite{es:Sud} for all those theories  related
to possible PEP violation  coming from new physics.

%

%
%




\begin{thebibliography}{99}




\bibitem{Pauli} W. Pauli, Phys. Rev. {\bf 58} (1940) 716.
\bibitem{Feynman} R. P. Feynman, R. B. Leighton, and 
M. Sands: "The Feynman Lectures on Physics", vol. 3, 
(Addison-Wesley, Reading, MA, 1963).
\bibitem{Bernabei} R. Bernabei {\it et al.}, Phys. Lett. {\bf B408} (1997) 439.
\bibitem{Borexino} H. O. Back {\it et al.} (Borexino Collaboration), Eur. Phys. J. {\bf C37} (2004) 421.
\bibitem{HH} R. C. Hilborn and C. L. Yuca, Phys. Rev. Lett. {\bf 76} (1996) 2844.
\bibitem{NEMO} NEMO Collaboration, Nucl. Phys. {\bf B87} (Proc. Suppl.) (2000) 510.
\bibitem{Nolte} E. Nolte {\it et al.}, J. Phys. G: Nucl. Part. Phys. {\bf 17} 
(1991) S355.
\bibitem{Tsi} Yu. M. Tsipenyuk, A. S. Barabash, V. N. Kornoukhov, and B. A. Chapyzhnikov, Radiat. Phys. Chem. {\bf 51} (1998) 507.
\bibitem{GM:Gre} O.W. Greenberg and R.N. Mohapatra, Phys. Rev. Lett. 
{\bf 59} (1987) 2507.
\bibitem{RS:RAM} E. Ramberg and G.A. Snow, Phys. Lett. {\bf B238} (1990) 
438.
\bibitem{VIPproposal} The VIP Proposal, LNF-LNGS Proposal, September 2004 

(http://www.lnf.infn.it/esperimenti/vip).
\bibitem{CCD} See e.g. J.L. Culhane, Nucl. Instr. and Meth. {\bf A310} (1991) 1; 
J.-P. Egger, D. Chatellard, E. Jeannet, Particle World {\bf 3} (1993) 139; 
G. Fiorucci {\it et al.}, Nucl. Instr. and Meth. {\bf A292} (1990) 141; 
D. Varidel {\it et al.}, Nucl. Instr. and Meth. {\bf A292} (1990) 147; 
R.P. Kraft {\it et al.}, Nucl. Instr. and Meth. {\bf A372} (1995) 372.
\bibitem{DEAR} T. Ishiwatari {\it et al.}, Phys. Lett. {\bf B593} (2004) 48;
G. Beer {\it et al.}, Phys. Rev. Let. {\bf 94} (2005) 212302.
\bibitem {PI:Pau} S. Di Matteo and L. Sperandio, VIP Note, IR-04, April 26, 2006
(the energy shift has been computed by P. Indelicato - private communication). 
\bibitem{xx:EEV} CCD-55 from EEV (English Electric Valve), Waterhouse Lane,
Chelmsford, Essex, CM1 2QU, UK.
\bibitem{NIM} T. Ishiwatari {\it et al.}, Nucl. Instrum. and Meth. in Phys.
Research {\bf A556} (2006) 509.
\bibitem{IK} A. Yu. Ignatiev and V. A. Kuzmin, Yad. Fiz. {\bf 46} (1987) 786; 
ICTP preprint IC/87/13 (1987); A. Yu Ignatiev, arXiv: hep-ph/0509258.
\bibitem{Govorkov} A. B. Govorkov, Phys. Lett {\bf A137} (1989) 7.
\bibitem{green} O.W. Greenberg, Phys. Rev. {\bf D43} (1991) 4111.
\bibitem{ok:Oku} L.B. Okun, Comments Nucl. Part. Phys. 19 (1989) 
998.
\bibitem{es:Sud} I. Duck and E.C.G. Sudarshan, Am. J. of Physics {\bf 66} (1998) 284.




\end{thebibliography}
\end{document}